# A Comparison of Estimand and Estimation Strategies for Clinical Trials in Early Parkinson's Disease

Alessandro Noci[1,*], Marcel Wolbers[1,*,+], Markus Abt[1], Corine Baayen[2], Hans Ulrich Burger[1], Man Jin[3], and Weining Zhao Robieson[3]

[1]Data and Statistical Sciences, Pharma Development, Roche, 4070 Basel, Switzerland
[2]Biometrics Division, H.Lundbeck A/S, 2500 Copenhagen, Denmark
[2]Data and Statistical Sciences, AbbVie Inc., North Chicago, IL 60064, USA
[*]These authors contributed equally to this work.
[+]Correspondence: marcel.wolbers@roche.com

**Abstract**

Parkinson's disease (PD) is a chronic, degenerative neurological disorder. PD cannot be prevented, slowed or cured as of today but highly effective symptomatic treatments are available. We consider relevant estimands and treatment effect estimators for randomized trials of a novel treatment which aims to slow down disease progression versus placebo in early, untreated PD. A commonly used endpoint in PD trials is the MDS-Unified Parkinson's Disease Rating Scale (MDS-UPDRS), which is longitudinally assessed at scheduled visits. The most important intercurrent events (ICEs) which affect the interpretation of the MDS-UPDRS are study treatment discontinuations and initiations of symptomatic treatment. Different estimand strategies are discussed and hypothetical or treatment policy strategies, respectively, for different types of ICEs seem most appropriate in this context. Several estimators based on multiple imputation which target these estimands are proposed and compared in terms of bias, mean-squared error, and power in a simulation study. The investigated estimators include methods based on a missing-at-random (MAR) assumption, with and without the inclusion of time-varying ICE-indicators, as well as reference-based imputation methods. Simulation parameters are motivated by data analyses of a cohort study from the Parkinson's Progression Markers Initiative (PPMI).

*Keywords:* Clinical Trials; Estimands; Missing Data; Multiple Imputation; Parkinson's disease; Reference-Based Methods

# 1 Introduction

Parkinson's disease (PD) is a chronic, degenerative neurological disorder characterized by movement problems such as rigidity, slowness, and tremor. More than 6 million individuals worldwide are affected by PD. At this time, no cure or disease-modifying therapy for PD exists. Levodopa preparations are the most widely prescribed treatments for PD. Levodopa is a highly effective symptomatic treatment for motor symptoms but it is not disease-modifying [Verschuur et al., 2019]. Moreover, over time, individuals require more frequent and higher levodopa doses and eventually lose their long-duration response to dopaminergic medication, and their short-duration response decreases. Other commonly used initial symptomatic treatments for PD include dopamine agonists and monoamine oxidase-B (MAO-B) inhibitors [Armstrong and Okun, 2020].

The most commonly used asssessment scale in clinical trials of PD is the Movement Disorder Society-sponsored revision of the Unified Parkinson's Disease Rating Scale (MDS-UPDRS) [Goetz et al., 2008]. The MDS-UPDRS is a multimodal scale assessing both impairment and disability. It consists of four subscales (Parts I-IV). Part I (13 items) measures non-motor experiences of daily living, Part II (13 items) assesses motor experiences of daily living, Part III (18 items with some items applying to multiple body parts resulting in 33 answers total) measures the motor signs of PD and Part IV (4 items) measures the motor complications, dyskinesias and motor fluctuations. Part IV is typically only assessed after a patient has started dopaminergic treatment. Each question is rated between 0-4, where 0 = Normal, 1 = Slight, 2 = Mild, 3= Moderate, 4 = Severe.

For the following, we assume that a double-blind, randomized trial comparing a new and potentially disease modifying intervention treatment versus placebo is to be conducted in early, untreated PD. The primary endpoint is the change from baseline to a fixed time point (e.g. at 12 months) in the MDS-UPDRS sum of Parts I+II+III score (called "MDS-UPDRS score" in the sequel for simplicity). The primary endpoint is assumed to be measured at regular intervals (e.g. every two months).

The aim of this article is to critically appraise several possible choices for the primary estimand in this setting, to propose treatment effect estimators which are aligned with these estimands, and to explore them in a simulation study. The article is structured as follows: Section 2 reviews the estimands framework and proposes three relevant estimands for our setting. Section 3 discusses corresponding treatment effect estimators. In section 4, the proposed estimands and estimators are compared in a simulation study. The simulation parameters are motivated by data analyses of a cohort study from the Parkinson's Progression Markers Initiative (PPMI) [Marek et al., 2011] which are reported in the supplementary Appendix. We conclude with a discussion of our findings and potential alternative estimands.

# 2 Estimands framework and definition of investigated estimands

The "ICH E9 (R1) addendum on estimands and sensitivity analysis in clinical trials to the guideline on statistical principles for clinical trials" presents a structured framework to link trial objectives to a precise description of the targeted treatment effect [ICH E9 working group, 2019]. The description of an estimand consists of five attributes: (A) the treatment, (B) the population, (C) the primary endpoint (or variable) to be obtained for each patient, that is required to address the scientific question, (D) the strategy of how to account for intercurrent events (ICEs) to reflect the scientific question of interest, and (E) the population-level summary for the variable which provides, as required, a basis for a comparison between treatment arms.

In our setting, the attributes of the primary estimands (apart from the ICEs which will be discussed afterwards) could be defined as follows:

(A) The treatment of interest is the randomized treatment (intervention treatment or matching placebo) administered as defined in the study protocol throughout the treatment period or until premature treatment discontinuation.

(B) The targeted population is an early, untreated PD population as defined by the trial inclusion/exclusion criteria.

(C) The primary endpoint is the change in the MDS-UPDRS (sum of Parts I+II+III) score from baseline to a fixed time point (e.g. at 12 months).

(D) The strategy of how to account for the ICEs is described in detail below.

(E) The population-level summary is the difference in the means of the primary endpoint between the randomized treatment groups.

ICEs are events occurring after treatment initiation that affect either the interpretation or the existence of the measurements associated with the clinical question of interest. The two most relevant ICEs which may occur in a trial in an early PD population are the discontinuation of the randomized study treatment and the initiation of symptomatic PD treatment with levodopa or other highly effective therapies. A patient may experience both of these ICEs in either order. In principle, study drug discontinuation could be categorized according to the discontinuation reason (e.g. as discontinued due to safety or tolerability reasons, due to lack of efficacy, or due to non study drug or condition related reasons) but we will not investigate this further here for simplicity reasons. Similarly, the ICE of death will not be discussed because it is expected to be rare in an early PD population.

For the purpose of the primary analysis, we consider treatment policy and hypothetical strategies to be most suitable to address the two ICEs of randomized treatment discontinuation and initiation of symptomatic treatment, respectively. Alternative strategies are revisited in the discussion section. For the treatment policy strategy, the occurrence of the ICE is considered irrelevant in defining the treatment effect of interest and the observed value for the endpoint is used regardless of whether or not the ICE occurs. In order to estimate a treatment effect under a treatment policy strategy without strong assumptions, it is critical that all patients are followed up to obtain their outcome assessments also after the ICE. For the hypothetical strategy, a scenario is envisaged in which the ICE would not occur. Under this strategy, compatible outcome values after the ICE are not directly observable and handled using models for missing data.

The ICH E9(R1) addendum explicitly allows that different strategies are applied to different ICEs and we consider three estimands. First, a pure hypothetical estimand which applies a hypothetical strategy to both study treatment discontinuations and initiations of symptomatic therapy. This envisions a hypothetical scenario where all patients fully adhere to the randomized study drug and symptomatic treatment is not available. It is acknowledged that it is impossible to envision a practical clinical trial where these ICEs would not occur. For example, it is the responsibility of the investigator based on benefit/risk considerations to withdraw a patient from study treatment in case the well-being of the patient might otherwise be compromised, and symptomatic medications must be made available for ethical reasons. However, a hypothetical estimand may still be relevant for phase 2 proof-of-concept trials. Second, we consider a mixed estimand which applies a treatment policy strategy to study treatment discontinuations and a hypothetical strategy to initiations of symptomatic therapy. This envisions a scenario where treatment discontinuations do occur and their impact on the outcome is of interest but where patients would not initiate symptomatic treatments (or, at least, defer symptomatic treatment until the end of follow-up). Third, we consider a pure treatment policy strategy where a treatment policy strategy is applied to both ICEs. An estimand which uses a hypothetical strategy for study treatment discontinuations and a treatment policy strategy for initiations of symptomatic treatment was not investigated because it was considered to be less clinically relevant and more artificial than the mixed estimand. All proposed estimands are summarizes in Table 1.

Table 1: Proposed estimands and corresponding strategies for the two ICEs.

|  | ICE | |
| --- | --- | --- |
|  | Study treatment discontinuation | Initiation of symptomatic treatment |
| Hypothetical estimand | Hypothetical strategy | Hypothetical strategy |
| Mixed estimand | Treatment policy strategy | Hypothetical strategy |
| Treatment policy estimand | Treatment policy strategy | Treatment policy strategy |

To illustrate the targeted estimands in the early PD setting, we assume that the intervention treatment slows down the rate of decline as assessed by the MDS-UPDRS score in the long term but, unlike the symptomatic treatment, it does not have a rapid and potent short-term effect on symptoms. This is visualized in Figure 1 which shows stylized MDS-UPDRS trajectories for two dummy subjects assuming that they had either been assigned to placebo or to intervention treatment, respectively. The first subject, represented by a solid black line, does not experience any ICEs and would have had a better outcome (represented by a more shallow slope) if assigned to the intervention. The second subject, represented by a solid gray line, would have discontinued randomized study treatment and subsequently initiated symptomatic treatment regardless of the assigned treatment group. If assigned to the placebo group, the subject's trajectory may be unaffected by study treatment discontinuation. However, initiation of symptomatic treatment may be associated with a rapid improvement in the MDS-UPDRS score. If assigned to the intervention group, the rate of disease progression may be slower initially but then revert to the placebo slope after treatment discontinuation. The second subject may also initiate symptomatic treatment in the intervention treatment group after their worsening in the MDS-UPDRS score from baseline exceeded 6 points and, again, they may subsequently experience a rapid improvement. The solid lines in Figure 1 correspond to the values relevant for the treatment policy estimand. In contrast, the dashed lines represent the relevant values for the mixed estimand which envisions a scenario without symptomatic treatment. It is important to note that, as shown in Figure 1, the pure treatment policy estimand favors patients who initiate symptomatic treatment. Indeed, the observed change in the MDS-UPDRS score at 12 months, which is relevant for the treatment policy estimand would assign a larger (i.e. worse) outcome to the first subject, represented by a black line, than to the second subject, represented by a gray line. However, it could be argued that the first subject has a more favorable outcome than the second subject. Thus the pure treatment policy estimand may implicitly rank patients in a counter-intuitive way in the presence of potent symptomatic treatment.

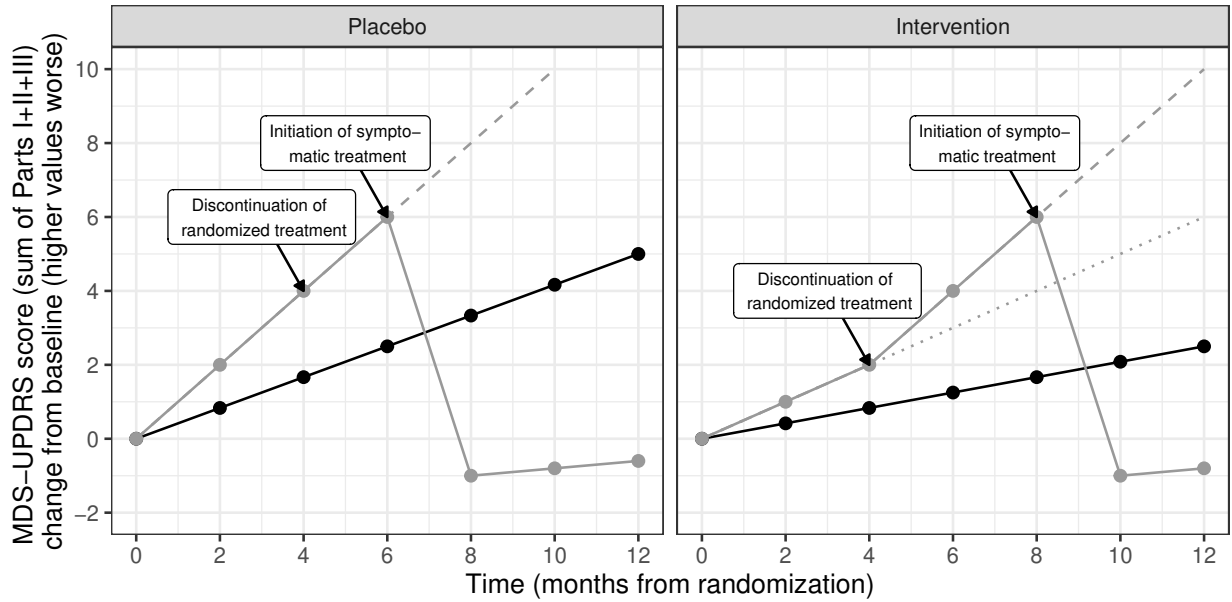

Figure 1: Stylized MDS-UPDRS trajectories for two (hypothetical) dummy subjects assuming that they had either been assigned to placebo or to intervention treatment, respectively. Black lines correspond to a subject who did not experience any ICEs. Grey lines correspond to a subject who discontinued from randomized treatment and subsequently initiated symptomatic treatment. Solid lines represent actual trajectories which are relevant for the treatment policy estimand. Dashed lines represent trajectories in the absence of symptomatic treatment which are relevant for the mixed estimand. The dotted lines represents a trajectory in the absence of randomized treatment discontinuation and symptomatic treatment which is relevant for the hypothetical estimand.

# 3 Missing data assumptions and estimators

## 3.1 Missing data assumptions and multiple imputation methods

It is critical to align the treatment effect estimator with the targeted estimand [Mallinckrodt et al., 2020]. In addition, the definition of a treatment effect estimator requires an appropriate handling of missing data in line with plausible assumptions about the underlying missing data mechanism. Missing data may occur due to missed assessments or study withdrawal. PD patients are fully aware of their disease and therefore interested in receiving the best possible care also after discontinuing study treatment. Therefore, we anticipate that in a carefully designed and executed study, the extent of missed assessments and study withdrawals will be relatively low. However, outcomes under a hypothetical strategy are not directly observable and any observed post-ICE outcomes are consequently ignored and treated as missing data under this strategy.

For the implementation of the estimators, we consider three different missingness mechanisms:

- *Baseline MAR*: Missing outcome data is similar to observed data from patients in the same treatment group with the same baseline characteristics and the same observed outcomes. More formally, this assumes that data are missing at random (MAR) after accounting for the randomized treatment group, baseline characteristics, and observed outcomes.

- *Reference-based missingness*: Until study treatment discontinuation, missing outcome data is similar to observed data from patients in the same treatment group with the same baseline characteristics who did not discontinue study treatment. After study treatment discontinuation, missing data is similar to observed data from patients in the placebo arm with similar baseline characteristics who did not discontinue study treatment. This assumes that treatment discontinuation does not impact outcomes in the placebo group but that patients randomized to the intervention treatment behave similarly to placebo group patients after discontinuation.

- *Time-dependent MAR*: Missing outcome data is similar to observed data from patients in the same treatment group with the same baseline characteristics who have the same ICE status at the relevant outcome visit. More formally, this assumes that data is MAR after accounting for the randomized treatment group, baseline characteristics, time-varying ICE indicators, and observed outcomes.

As in Polverejan and Dragalin [2020], we use multiple imputation (MI) methods for treatment effect estimation because this provides a flexible framework for imputing under different missingness mechanisms and different imputation techniques for different categories of subjects. We focus on conventional Bayesian MI as described in Little and Rubin [2002] and Carpenter et al. [2013] but acknowledge that implementations based on maximum likelihood imputation and re-sampling have also been proposed [von Hippel and Bartlett, 2021, Wolbers et al. [2021]].

In our context, Bayesian MI consists of the following steps. First, a Bayesian imputation model is fitted which is a multivariate normal Bayesian mixed model for repeated measures (MMRM) of the outcomes depending on the randomized treatment group, the visit, visit-by-treatment interactions, and appropriate baseline covariates. Frequently, a common unstructured covariance matrix is assumed for both treatment groups. For models assuming time-dependent MAR, additional time-varying covariates are included in the imputation model. If post-discontinuation outcomes are available in the data set, MI under a reference-based missingness assumption is implemented by excluding post-discontinuation outcomes from the imputation model fit but keeping them in the subsequent imputation and analysis steps as described in Wolbers et al. [2021]. Second, for each posterior parameter draw from the imputation model, an imputed data set is generated. This proceeds by randomly imputing missing data conditional on the patient's baseline characteristics and observed outcomes, and the patient's marginal mean and covariance matrix predicted by the parameters from the imputation model. Under a reference-based missingness assumption, the marginal imputation mean appropriately combines mean trajectories from the assigned arm and the placebo arm after discontinuation as described in Carpenter et al. [2013, section 4.3]. Third, each imputed data set is analyzed. As imputed data sets are complete, the analysis step may consist of fitting a simple ANCOVA model to the change in the MDS-UPDRS score from baseline to a fixed time point (e.g. at 12 months) with the treatment group as the main covariate and adjustment for the baseline MDS-UPDRS score and other important baseline covariates. Finally, inferences are pooled across multiple imputed data set using Rubin's rules. For a more technical and complete discussions of Bayesian MI methods we refer to Little and Rubin [2002] and Carpenter et al. [2013].

## 3.2 Estimators for the hypothetical estimand

For the hypothetical estimand, any outcomes collected after treatment discontinuation or initiation of symptomatic treatment are not of interest and considered as missing for the purpose of treatment effect estimation. The baseline MAR assumption is often a good starting point for implementing a hypothetical strategy [Mallinckrodt et al., 2020] and can be implemented as described in the previous section. In the sequel, we refer to this estimator as *MAR(hypothetical)*. Of note, in this setting, the MI-based estimator is asymptotically equivalent to the popular MMRM model [Mallinckrod et al., 2008, Siddiqui [2011]].

## 3.3 Estimators for the mixed estimand

The mixed estimand applies a treatment policy strategy to study treatment discontinuations and a hypothetical strategy to symptomatic treatment initiations. Therefore, outcomes collected after symptomatic treatment initiations are not of interest and considered as missing for the purpose of treatment effect estimation. In contrast, outcomes collected after study treatment discontinuation but prior to symptomatic treatment are compatible with the treatment policy strategy and will be included in the analysis.

A patient may experience both study treatment discontinuations and symptomatic treatment initiations in either order. A complication for the estimator of the mixed estimand arises for patients who initiate symptomatic treatment prior to study treatment discontinuation. The reason is that for these patients it is unknown whether and when they would have subsequently discontinued study treatment in the (hypothetical) absence of symptomatic treatment discontinuation. For our estimators, we assume for simplicity reasons that the likelihood of study treatment discontinuations does not depend on whether the patient previously initiated symptomatic medication or not, i.e. we base estimation on the observed ICEs of treatment discontinuation. Without such an assumption, estimation would be even more involved and require modeling or imputing the ICE study treatment discontinuation in the absence of observed symptomatic treatment discontinuations. As only few patients are expected to experience multiple ICEs in a trial, we believe that such a simplification is justifiable.

Four different estimators for the mixed estimands will be explored in the simulation study and are described below: one estimator assuming baseline-MAR, one estimator assuming reference-based missingness, and two estimators assuming time-dependent MAR.

The estimator based on a baseline-MAR assumption is referred to as *MAR(mixed)* in the sequel. Importantly, imputations based on the baseline-MAR assumption do not account for the effect of treatment discontinuation whereas, typically, patients who discontinue active treatment would be expected to benefit less (or not at all) from the randomized treatment. Moreover, missing data is much more likely to occur after treatment discontinuation. In practice, this implies that the imputation model which is based on observed data is dominated by on-treatment outcomes and, consequently, imputations after treatment discontinuation may be over-optimistic.

Reference-based imputation models were introduced by Carpenter et al. [2013] to formalize the idea of imputing missing data after treatment discontinuation in both arms based on the control arm. A popular reference-based imputation method is copy-increments in reference (CIR) which will be used for the estimator *CIR(mixed)*. In the placebo control arm, CIR assumes no impact of treatment discontinuations whereas in the active arm, it assumes that a patient's post-discontinuation mean increments are equal to those from the placebo arm. Informally, this means that the treatment benefit accrued up to discontinuation is retained but that there is no additional residual benefit after discontinuation which may be considered a relatively conservative imputation strategy for an intervention which aims to slow down disease progression.

Estimators based on a time-dependent MAR assumption have been proposed by Guizzaro et al. [2021]. Rather than building separate imputation models for pre- and post-discontinuation data, we assume that discontinuation only affects a patient's mean trajectory and that this can be modeled by including appropriate time-varying covariates. In principle, the effect of treatment discontinuation on outcomes may depend on both the timing of the discontinuation and the time between the discontinuation and the visit at which the outcome value is to be imputed, but in practice more parsimonious models with fewer degrees of freedom are required. Specifically, we model the effect of treatment discontinuation either by including treatment arm specific binary time-varying post-discontinuation indicators (estimator *TV1-MAR(mixed)*) or time-varying variables which are set equal to 0 up to the treatment discontinuation and to the time from treatment discontinuation at subsequent visits (estimator *TV2-MAR(mixed)*). The former model assumes that the effect of treatment discontinuation is a constant shift in outcomes whereas the latter implies a change in the slope of outcome trajectories. The latter may be more plausible for a disease-modifying treatment in early PD given that changes in MDS-UPDRS scores over time are approximately linear in untreated patients as shown in the supplementary Appendix.

### 3.4 Estimators for the treatment policy estimand

All measured outcomes before or after any ICE are relevant for a pure treatment policy estimand and are therefore included in the analysis. Ideally, no missed outcome assessment or study withdrawals occur so that no imputation would be necessary for the treatment policy estimand. In the presence of missing data, we investigate three different estimators: a simple estimator based on the baseline-MAR assumption (*MAR(treatment policy)*) and two estimators which include time-varying covariates (*TV3-MAR(treatment policy)* and *TV4-MAR(treatment policy)*). The two estimators under a time-dependent MAR assumption use the same time-varying covariates to model post-discontinuation data as described for the mixed estimand. In addition, both models include treatment group specific binary time-varying indicators of initiation of symptomatic treatment.

All examined estimators are summarized in Table 2.

Table 2: Proposed estimators and imputation strategies for all estimands.

|  | Imputation strategy |
|---|---|
| MAR (hypothetical)[a] | Impute based on baseline characteristics and assigned treatment group. |
| MAR (mixed)[b] | Impute based on baseline characteristics and assigned treatment group. |
| CIR (mixed)[b] | As for MAR (hypothetical) prior to study treatment discontinuation, according to increments in control group after discontinution. |
| TV1-MAR (mixed)[b] | Impute based on baseline characteristics, assigned treatment group, and a time-varying binary 'post treatment discontinuation' indicator. |
| TV2-MAR (mixed)[b] | Impute based on baseline characteristics, assigned treatment group, and a time-varying 'time since treatment discontinuation' variable. |
| MAR (treatment policy)[c] | Impute based on baseline characteristics and assigned treatment group. |
| TV3-MAR (treatment policy)[c] | As for TV1-MAR but include also a time-varying binary 'on symptomatic treatment' indicator. |
| TV4-MAR (treatment policy)[c] | As for TV2-MAR but include also a time-varying binary 'on symptomatic treatment' indicator. |

[a] Estimators for the hypothetical estimand include only pre-ICE data.
[b] Estimators for the mixed estimand include only data before symptomatic treatment initiation.
[c] Estimators for the treatment policy estimand include all observed outcome data.

## 4 Simulation study

### 4.1 Data simulation

Parameters for modeling MDS-UPDRS trajectories in the control group, initiations of symptomatic treatment, and the effect of symptomatic treatment on MDS-UPDRS trajectories were motivated by detailed data analyses of the Parkinson's Progression Markers Initiative (PPMI) database. Longitudinal changes in the MDS-UPDRS score and other clinical and biological measures were published by Simuni et al. [2018]. Analyses which are targeted to our simulations and use a more recent cut-off of the PPMI database are summarized in the supplementary Appendix. Horvath and colleagues estimated clinically meaningful within-patient changes in MDS-UPDRS scores using anchor-based analyses and determined thresholds of 2-3 points for the MDS-UPDRS part I and II scores and 3-5 points for the MDS-UPDRS part III score, respectively [Horvath et al., 2015, Horvath et al. [2017]]. Acknowledging that a new treatment may not affect all MDS-UPDRS parts equally, we chose a target treatment effect of -4 points (+6 points/year on intervention vs +10 points/year on placebo) for the hypothetical estimand in our simulations. In order to pressure test the different estimators, we simulated study treatment discontinuation rates and study drop-out rates that are at the upper range of what might be expected in an actual trial in early PD.

Specifically, the simulation study generated hypothetical 1:1 randomized trials with the following parameters:

- The sample size was varied from 75 to 300 patients per group in increments of 25 patients.
- The treatment duration of the trials was 12 months with bi-monthly visit from randomization until 12 months.
- The mean trajectory of the MDS-UPDRS score in the placebo group in the absence of ICEs increased linearly by 10 points from 30 points at baseline to 40 points at 12 months.
- The mean trajectory of the MDS-UPDRS score in the intervention group in the absence of ICEs increased linearly by 6 points from 30 points at baseline to 36 points at 12 months.
- The covariance structure of the baseline and follow-up values in both groups was implied by a random intercept and slope model with a standard deviation of 10 for the intercept and of 5 (per year) for the slope, and a correlation of 0.5. In addition, an independent residual error with standard deviation 6 was added to each assessment. This implies marginal standard deviations ranging from 11.7 at baseline to 14.5 at the 12 month visit in both groups.
- The probability of study treatment discontinuations after each visit was set to 2% for the placebo group (implying an overall discontinuation probability of 11.4% before the 12 month visit) and to 3% after each visit for the intervention group (overall 16.7%). A slightly higher discontinuation rate in the treatment arm is plausible in settings where treatment discontinuation are primarily due to safety and tolerability issues and not due to lack of efficacy.
- Study treatment discontinuations were simulated to have no impact on MDS-UPDRS trajectories in the placebo group. For the intervention group, it was assumed that the fixed slope increased to the placebo group slope after discontinuation.
- Patients who discontinued study treatment had a 50% probability of withdrawing from the study leading to missing outcome data from the time of discontinuation onward.
- The probability of initiation of symptomatic treatment after each visit was simulated independently of study treatment discontinuation. In line with analyses of the PPMI database reported in the supplementary Appendix, the probability was dependent on the observed MDS-UPDRS score with an 1.5-fold higher odds for each +10 points increase. That is, if the MDS-UPDRS score at a visit differed by 10 points between two subjects, then the subject with the higher score had a 1.5-fold higher odds of initiating symptomatic treatment at that visit. Moreover, a lower probability of symptomatic treatment initiation was chosen for earlier visits. Specifically, for a patient with an MDS-UPDRS score of 30 at a visit, the probability was 0% to discontinue after the baseline visit, 2.5% after the month 2 and 4 visits, and 7.5% after subsequent visits. This implies that the overall probability of symptomatic treatment initiation during the 12 month study was 33% in the placebo group and 30% in the intervention group.
- Initiation of symptomatic treatment was associated with an immediate improvement in the MDS-UPDRS score according to a re-scaled beta-distribution with parameters $\alpha = 2$, $\beta = 1.5$ and range from -25 to 0 points. This implies a median drop by -10.34 points (interquartile range -15.14 to -5.97 points). Moreover, the fixed slope after initiation of symptomatic treatment was set to 0. It is worth remarking that an almost flat slope was estimated from the PPMI data by considering only the first 12 months of follow-up (see the supplementary Appendix for details). This does not assume that patients won't have a worsening of the disease later in time.

The targeted treatment effects by the three estimands, i.e. the simulation truth, was determined by simulating a large trial with $n = 3,000,000$ subjects per treatment group and were -4.00 points (+6.00 vs +10.00) for the hypothetical estimand, -3.60 points (+6.40 vs +10.00) for the mixed estimand, and -2.85 points (+2.36 vs +5.21) for the treatment policy estimand. Marginal mean trajectories in both treatment groups corresponding to these estimands are shown in Figure 2 and illustrate the pronounced impact of symptomatic treatment initiations on the treatment policy estimand.

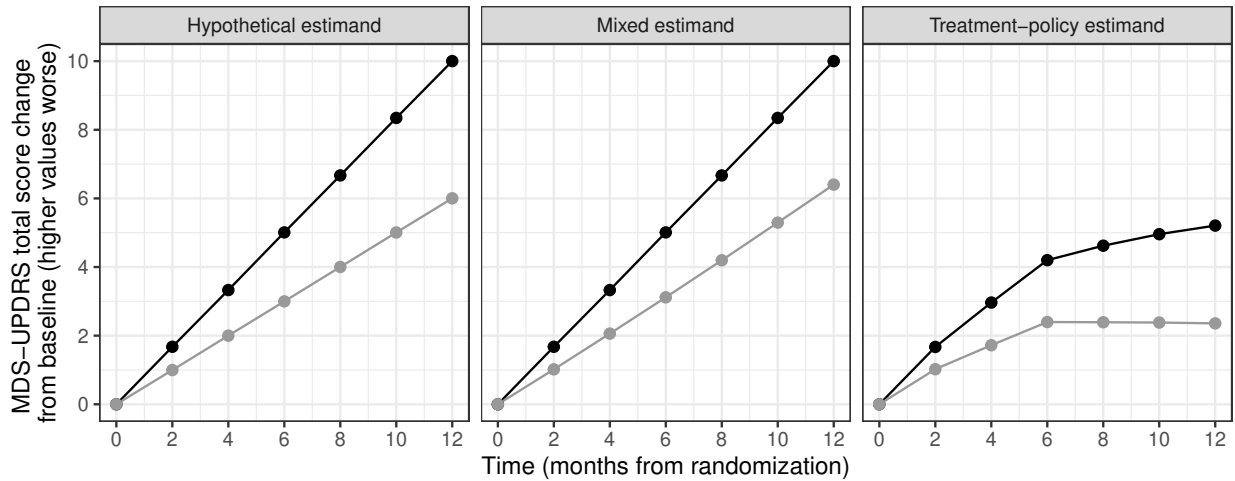

Figure 2: Marginal mean MDS-UPDRS trajectories for the intervention (gray lines) and placebo (black lines) groups in the simulation study relevant to the hypothetical, mixed, and treatment policy estimands, respectively. The targeted treatment effect ('simulation truth') at 12 months is -4.00 points for the hypothetical estimand, -3.60 for the mixed estimand, and -2.85 for the treatment policy estimand.

### 4.2 Implementation of estimators and other simulation study parameters

Estimators were implemented as described in Section 3. Specifically, we used Bayesian multiple imputation (with $M = 100$ random imputations per data set) and Rubin's rules for pooling inferences across multiple imputed data sets. The imputation model had the mean change from baseline in the MDS-UPDRS score as the outcome, the treatment group, the (categorical) visit, treatment-by-visit interactions, the baseline MDS-UPDRS score, and baseline MDS-UPDRS score-by-visit interactions as covariates, and assumed a common unstructured covariance matrix in both groups. For estimators involving time-varying covariates, these covariates (as well as treatment by time-varying covariate interaction terms) were additionally included in the imputation model. The analysis model was an ANCOVA model with the mean change from baseline in the MDS-UPDRS score at 12 months as the outcome, the treatment assignment as the main covariate, and adjustment for the baseline MDS-UPDRS score. For each scenario and estimator, results were averaged over 10,000 simulated data sets. Reported performance measures are the bias and mean-squared error for each treatment effect estimator and the power of the associated significance test of the null hypothesis of a treatment effect of 0 at a two-sided significance level of 5%.

### 4.3 Simulation results

The performance of all estimators for the three estimands is summarized in Table 3. Only the results for the scenarios with a sample size of 300 patients per group are reported. The reason for this is that the empirical bias did not materially depend on the sample size. Moreover, the root mean-squared error (RMSE) of the estimates and the average standard error associated with the estimates decreased with increasing sample size as expected. The simulated power of the estimators

Table 3: Performance of all estimators for the three estimands for a sample size of 300 patients per treatment group based on 10,000 simulated datasets.

|  |  | Estimate | | | |
| --- | --- | --- | --- | --- | --- |
| Estimator | Targeted effect ('truth') | Mean | Bias | RMSE | Mean SE |
| MAR (hypothetical) | -4.00 | -3.99 | 0.01 | 0.95 | 0.95 |
| MAR (mixed) | -3.60 | -3.77 | -0.18 | 0.94 | 0.92 |
| CIR (mixed) | -3.60 | -3.59 | 0.01 | 0.86 | 0.91 |
| TV1-MAR (mixed) | -3.60 | -3.68 | -0.08 | 0.94 | 0.93 |
| TV2-MAR (mixed) | -3.60 | -3.59 | 0.01 | 0.94 | 0.94 |
| MAR (treatment policy) | -2.85 | -2.97 | -0.12 | 0.97 | 0.98 |
| TV3-MAR (treatment policy) | -2.85 | -2.89 | -0.04 | 0.97 | 0.98 |
| TV4-MAR (treatment policy) | -2.85 | -2.81 | 0.04 | 0.98 | 0.99 |

RMSE = empirical root mean-squared error of estimates, mean SE= mean of the standard errors associated with the 10,000 treatment effect estimates.

for all investigated sample sizes is presented in Figure 3.

For the hypothetical estimand, the baseline-MAR estimator based only on pre-ICEs data had negligible empirical bias. It was also associated with the largest study power compared to all other estimators. Compared to the CIR estimator for the mixed estimand and estimator TV4-MAR with time-varying covariates for the treatment policy estimand, the two estimators with lowest bias for these two estimands, its power was 66% compared to 62% and 37%, respectively, for a sample size of 100 patients per group, and 99% compared to 98% and 81% for a sample size of 300.

For the mixed estimand, the estimator based on CIR imputation and estimator TV2-MAR based on a time-varying covariate representing the time from treatment discontinuation had negligible bias. Estimator TV1-MAR which included a time-varying indicator of treatment discontinuation slightly overestimated the magnitude of the treatment effect. This bias is due to a misspecification of the time-varying effect of treatment discontinuation on outcomes compared to the simulation parameters. The baseline-MAR estimator more substantially overestimated the magnitude of the treatment effect because the imputation did not take into account treatment discontinuations. The RMSE was comparable for all MAR-based estimators but substantially lower for the CIR-based estimator. This is well known and occurs because reference-based imputations in the intervention group borrow information from the control group which induces a positive correlation between the two groups and a corresponding reduction in the frequentist variance of the treatment effect contrast [Seaman et al., 2014, Wolbers et al. [2021]]. This variance reduction is not properly captured by Rubin's rules which explains why the average standard error of the treatment effect estimates is only slightly reduced compared to the baseline-MAR estimator [Bartlett, 2021]. Estimators with time-varying covariates had slightly elevated average standard errors. In terms of power, the baseline MAR estimator had the highest power, followed by the CIR-based estimator, and estimators TV1-MAR and TV2-MAR based on a time-dependent MAR assumption. The corresponding power estimates were 64%, 62%, 60%, and 57% for a sample size of 100 and 91%, 91%, 89%, and 86% for a sample size of 200 patients per group. The slightly larger power for the baseline MAR estimator over the CIR-based estimator is likely an artifact of the anti-conservative bias of the baseline MAR estimator.

For the treatment policy estimand, the baseline-MAR estimator based on all observed data had the largest bias and over-estimated the magnitude of the treatment effect. The estimators based on a time-dependent MAR assumption reduced the bias by a factor of three but still had a small bias because the specified time-varying indicator variable representing the impact of symptomatic treatment on the outcome was not fully compatible with the simulated effect. The targeted treatment effect of the treatment policy estimand was substantially smaller than for the other estimands and the variability of the estimators (represented by the mean standard error) was substantially larger because of the increased variability in the longitudinal data caused by the symptomatic treatment. Consequently, a larger sample size of 300 patients per group was required until all estimators were associated with a study power above 80% (86% for the baseline-MAR estimator, 84% for estimator TV3-MAR, and 81% for estimator TV4-MAR).

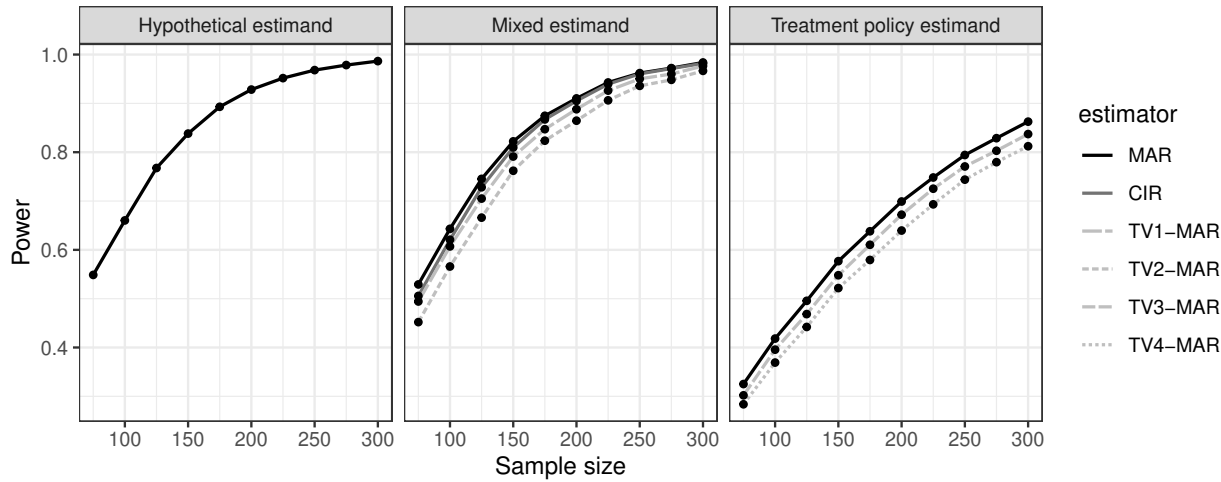

Figure 3: Estimated power of all estimators for the three estimands based on 10,000 simulated datasets per sample size.

A last remark regarding the number of failures to obtain a valid treatment effect estimate based on MI. Our Bayesian MI implementation relied on an initial frequentist MMRM fit of the base imputation model with an unstructured covariance matrix to inform the starting values and the prior distribution for the Bayesian imputation model. This is in accordance with the SAS implementation of reference-based imputation by the Drug Information Association Scientific Working Group on Estimands and Missing Data [Roger, 2021]. If the initial MMRM model failed to converge, then no treatment effect estimator could be determined for a simulation run and it was consequently omitted from all summaries reported above. For all estimators without time-varying covariates, convergence failures were extremely rare and the maximum number of failures observed for any estimator and sample size was 0.06% (6/10,000). For estimators TV1-MAR and TV2-MAR, we observed 2.50% (250/10,000) and 2.49% (249/10,000) failures for a sample size of 75 patients per group, respectively, and 0.73% (73/10,000) failures for both estimators for a sample size of 100 per group. For larger sample sizes, the number of failures for these estimators never exceeded 0.20% (20/10,000). For estimators TV3-MAR and TV4-MAR, we observed 1.44% (144/10,000) and 1.46% (146/10,000) failures for a sample size of 75 patients per group, respectively, and 0.29% (29/10,000) failures for both estimators for a sample size of 100 per group. For larger sample sizes, the number of failures for these estimators never exceeded 0.12% (12/10,000). The larger proportion of failures of estimators involving time-varying covariates occurs because they fail if no no data is available to inform the regression parameter of the time-varying covariates. Our simulation set-up implies that the probability that a randomly chosen patient discontinues study medication and has at least one observed post-discontinuation outcome prior to symptomatic treatment initiation is 4.98% in the control group and 7.36% in the intervention group. For simulated dataset with a sample size of 75 patients per group, this implies a probability of 2.48% that no post-discontinuation outcomes prior to symptomatic treatment initiation are included in at least one of the treatment groups. The probability of 2.48% is almost identical to the reported proportion of failures for the estimators TV1-MAR and TV2-MAR of the mixed estimand for this sample size.

# 5 Discussion

We proposed three different estimands and corresponding treatment effect estimators for an intervention treatment which aims to slow down disease progression in early PD and compared them in a realistic simulation study. A key challenge in this setting is that a substantial proportion of patients may initiate potent symptomatic treatment which could mask the effect of an intervention with significant long-term benefits but less potency for short-term symptom relief. None of the three estimands is uniformly superior to the others. The treatment effect of the hypothetical estimand is relatively straightforward to estimate using a standard MMRM model or multiple imputation under a baseline MAR assumption. The hypothetical estimand was also associated with the highest power in our simulation study. However, the clinical relevance of the hypothetical estimand is not clear because it is impossible to envision a practical scenario where neither treatment discontinuations nor initiations of symptomatic treatment would occur. Therefore, we believe that the hypothetical estimand is primarily relevant to signal-seeking and dose-finding trials. The mixed estimand corresponds to a somewhat more realistic scenario where treatment discontinuations are possible but symptomatic treatments are not allowed (or, more pragmatically, would be deferred until the end of the one year follow-up). For treatments in Alzheimer's disease dementia, the guidance document by the European Medicines Agency states that "an appropriate target of estimation could be based on a hypothetical scenario in which the new concomitant medication or modifications in the dose of concomitant medications had not been introduced" [Committee for Medicinal Products for Human Use (CHMP), 2018]. Thus, a hypothetical strategy for symptomatic treatment initiations might also be acceptable to regulators in early PD. The treatment policy estimand is closest to the traditional intention-to-treat principle but implicitly favors patients who initiate symptomatic medications. It is also associated with a substantially attenuated treatment effect and larger variability. A consequence of this is that trials powered for the treatment policy estimand would require substantially larger sample sizes than the other estimands ($n \approx 300$ patients per group compared to $n \approx 175$ for the mixed and $n \approx 150$ for the hypothetical estimand, respectively, for $>80\%$ power in our simulation study). In our simulation study, the probability of symptomatic treatment initiation during the 12 month study was 33% in the placebo group and 30% in the intervention group, i.e. it was only marginally higher in the placebo group. If the difference in the rate of symptomatic treatment initiations between treatment groups was substantially larger, e.g. due to a very large intervention effect, then this could, at worst, completely mask a relevant treatment effect for the treatment policy estimand.

The discussed missing data assumptions and estimators illustrate the complexity of handling multiple ICEs which may occur in either order and may be addressed by different strategies. For the mixed and the treatment policy estimand, a baseline MAR assumption for missing data is no longer tenable and the associated estimators had the largest (and anti-conservative) bias. Alternatives based on reference-based imputation or MAR imputation based on models which include time-varying covariates eliminated the bias or reduced it substantially. For a treatment which aims to slow down disease progression, the reference-based CIR assumption is quite plausible and relatively conservative. The associated treatment effect estimator was unbiased and had slightly larger power than imputations with time-varying covariates for the mixed estimand in our simulation study. Of note, the power gain of the CIR estimator is expected to be even larger if inference was based on frequentist-valid standard errors rather than the conventional Rubin's rules [Seaman et al., 2014, Bartlett [2021],Wolbers et al. [2021]]. A disadvantage of reference-based imputation methods is that imputations are not informed by retrieved post-discontinuation data. If post-discontinuation data is frequently collected, then models including time-varying covariates are a less assumption-dependent alternative [Guizzaro et al., 2021]. In our simulation study, these methods were successful at reducing bias and had acceptable power despite a relatively large simulated study drop-out rate of 50% after treatment discontinuation. However, these methods are also no panacea because they rely on realistic models for post-discontinuation outcomes.

All proposed estimators rely on missing data assumptions and it is important to pressure test the impact of these assumptions via suitable sensitivity analyses. As an example, the MAR assumption for the hypothetical estimand would be violated if longitudinal assessments other than the MDS-UPDRS score existed which affected both the occurrence of the ICEs and also subsequent MDS-UPDRS scores. A further complication occurs if all patients whose MDS-UPDRS score exceeded a certain threshold deterministically discontinued study treatment or initiated symptomatic treatment. This would lead to a violation of the so-called positivity assumption. As a consequence, methods based on MMRM or MI are still feasible, but the validity of their estimates relies on extrapolation beyond the data [Parra et al., 2021]. The positivity assumption is unlikely to be violated in the early PD setting. Specifically, analysis of the PPMI cohort reported in the supplementary Appendix demonstrates that while the probability of symptomatic treatment initiation depends on the MDS-UPDRS score, the association is far from deterministic. However, it remains plausible that patients who experience an ICE may have had worse outcomes than predicted by MAR if the ICE could have been prevented. One possibility to stress-test the impact of the MAR assumption on treatment effect estimates and inference are sensitivity analyses which apply a $\delta$-adjustment to MAR-imputed data prior to the analysis to reflect the anticipated magnitude of this worsening [Cro et al., 2020].

Alternative strategies for dealing with the ICE of symptomatic treatment initiations may also be considered. Darken et al. [2020] proposed a so-called "attributable estimand" which would handle the ICE with a composite strategy. In practice, this can be implemented by imputing suitably unfavorable outcomes for patients with the ICE. However, it is difficult to justify the exact penalty that can be considered as "suitably unfavorable". The principal stratification strategy would determine treatment effects in the subgroup of patients who would not have initiated symptomatic treatment regardless of the assigned treatment group [Bornkamp et al., 2021]. Effects in principal strata can only be estimated under assumptions that cannot be tested empirically and the clinical relevance of such an estimand is less clear. Finally, a while on treatment strategy could be applied and the last outcome value prior to symptomatic treatment initiation rather than the outcome value at a fixed time point could be used in the analysis. Again, the clinical relevance of such a strategy is less clear. Therefore, we believe that these alternative strategies are usually better suited for exploratory and sensitivity analyses than for a primary analysis.

Another possibility is to change the endpoint and the outcome measure altogether. For disease-modifying treatments, a slope model which assumes a linear progression rate has been proposed as an alternative to the MMRM model. A comparison of slopes between treatment groups is typically more powerful than a comparisons of changes at a fixed time point but it also relies on stronger assumptions [Chen et al., 2018]. As another possibility, a time-to-event endpoint such as the time to an increase in the MDS-UPDRS score by a pre-determined threshold could be defined. Many patients may reach this endpoint prior to the initiation of symptomatic treatment hence limiting its impact. It would be interesting to compare such alternative approaches to the endpoints discussed in this article. Finally, the time to initiation of symptomatic could also be considered as an endpoint. Our analyses of the PPMI data showed a clear association between higher MDS-UPDRS scores and the likelihood of symptomatic treatment initiation. However, the association was far from deterministic and it seems likely that subjective factors such as doctor and patient preferences also affect the timing of symptomatic treatment initiation.

A general limitation of clinical trials with a limited follow-up duration of 12-24 months and the proposed MDS-UPDRS-based estimands is that they can only demonstrate a relatively short-term delay or slowing of disease progression. As stated in the guideline on clinical investigation of medicinal products in the treatment of PD by the European Medicines Agency, this does not imply that a new agent is also a disease modifier. The latter would require the additional demonstration of an effect on the underlying pathophysiology of the disease by e.g. biochemical markers or neuroimaging measures [Committee for Medicinal Products for Human Use (CHMP), 2012]. Moreover, long-term follow-up of patients after the primary endpoint assessment is desirable.

In conclusion, we discussed the complexity of defining suitable estimands and estimators in early PD and suggested three estimands and corresponding treatment effect estimators. Estimators based on reference-based imputation or imputation models which included time-varying covariates improved upon standard MI or MMRM analyses which had substantially larger bias. More research and regulatory guidance is needed to establish consensus endpoints and estimands for developing novel treatments in early PD.

## Software implementation

R code for the reported simulation study has been uploaded to github and can be found at github.com/nociale/Simulation_PDestimands_manuscript. The implementation of all multiple imputation estimators is based on the R package 'rbmi' (Reference Based Multiple Imputation) which will be uploaded to CRAN in early 2022.

## Acknowledgements

We thank Annabelle Monnet and Judith Anzures-Cabrera from Roche and Khadija Rantell and Sabine Lenton from the UK Medicines and Healthcare products Regulatory Agency (MHRA) for helpful comments on an earlier draft of the manuscript which helped to improve the final presentation.

As described, simulation parameters were motivated by analyses of data obtained from the Parkinson's Progression Markers Initiative (PPMI) database (www.ppmi-info.org/access-dataspecimens/download-data). For up-to-date information on the PPMI study, visit ppmi-info.org. PPMI – a public-private partnership – is funded by the Michael J. Fox Foundation for Parkinson's Research and funding partners listed at www.ppmi-info.org/about-ppmi/who-we-are/study-sponsors.

# Appendix to the Manuscript 'A Comparison of Estimand and Estimation Strategies for Clinical Trials in Early Parkinson's Disease': Analysis of Data from the Parkinson's Progression Markers Initiative (PPMI) to Motivate the Simulation Study

## 1 Introduction

This appendix reports analyses of data from the Parkinson's Progression Markers Initiative (PPMI) which were performed to inform the simulation study of the manuscript "A comparison of estimand and estimation strategies for clinical trials in early Parkinson's disease".

## 2 Methods

### 2.1 Data

Data were obtained from PPMI (https://www.ppmi-info.org/, download of data on 30Nov2020). The population consisted of all subjects in the PPMI Parkinson's disease (PD) cohort who were 40-80 years old, had their PD diagnosis less than 2 years prior to enrolment, had a baseline Hoehn and Yahr scale of 1 or 2, a non-missing assessment of the MDS-UPDRS score (sum of parts I-III), and were not on any PD medication at baseline, and not expected to require PD medication within at least 6 months from baseline. These criteria mimic plausible enrolment criteria for a RCT in early PD.

The data analysis included the scheduled assessments at baseline and 3, 6, 9, and 12 months of follow-up and any other visits during the first year (defined as any visit occurring within 13.5 months of baseline to accomodate delayed month 12 visits). Symptomatic dopaminergic treatment was defined as any treatment with levadopa or dopamine agonists. The date of initiation of symptomatic dopaminergic treatment was defined as the mid-date between the last visit off and the first visit on dopaminergic treatment.

The primary outcome was the Unified Parkinson's Disease Rating Scale (MDS-UPDRS) sum of Parts I+II+III score (called "MDS-UPDRS score" in the sequel for simplicity). For patients on dopaminergic treatment, the part III assessment in a practically defined "on" medication state was used for subjects, i.e. assessments were taken within <6 hours of treatment. The PPMI dataset also contains MDS-UPDRS part III assessment in a practically defined "off" state ($\geq 6$ hours after the last dose of dopaminergic treatment) but they were only scheduled annually for subjects on dopaminergic treatment. Our preference for using "on" medication assessments has three reasons: First, such measurements are more likely to reflect the state of subjects in clinical practice where patients would typically be kept in "on" state. Second, the definition of the "off" state is tailored to levodopa treatment with its short half-life whereas dopamine agonists typically have a much longer half life. Third, the less frequent assessments in "off" state would complicate longitudinal modeling.

## 2.2 Statistical methods

Longitudinal MDS-UPDRS scores during the first year of follow-up and prior to initiation of symptomatic dopaminergic treatment were modeled using linear mixed effect models with random (subject-specific) intercept and slope terms. Longitudinal MDS-UPDRS scores after initiation of symptomatic dopaminergic treatment were modeled with linear mixed effect models with a random intercept and a fixed slope. A random slope term could not be fitted to this latter model because insufficient MDS-UPDRS data after dopaminergic treatment initiation and within 12 months of baseline was available.

The probability of initiation of symptomatic dopaminergic treatment after the baseline, month 3, month 6, or month 9 visit, respectively, was modeled using a logistic regression model with the visit and the current MDS-UPDRS score as covariates. It was also investigated whether the change in the MDS-UPDRS score from the previous visit was a predictor for treatment initiation. The analysis was based on stacked data including all visits of a subject prior to initiation of dopaminergic treatment. To accommodate that a subject may contribute multiple rows to this dataset, estimation was based on generalized estimating equations (GEE) with working independence correlations structure and corresponding robust standard errors.

All analyses were conducted with the statistical software `R v4.0.3` and its companion packages `lme4` and `geepack`.

# 3 Results and implications for the simulation study

## 3.1 Baseline characteristics

Table 1 summarizes the characteristics of the included 401 subjects.

Table 1: Baseline characteristics of included subjects.

|  | Overall (N=401) |
|---|---:|
| **Age [years]** | |
|   N | 401 |
|   Mean (SD) | 62.12 (8.61) |
|   Median (Q1, Q3) | 63.00 (56.00, 69.00) |
|   Range | 40.00 - 80.00 |
| **Sex** | |
|   female | 133 (33.2%) |
|   male | 268 (66.8%) |
| **Time since PD diagnosis [years]** | |
|   N | 401 |
|   Mean (SD) | 0.42 (0.46) |
|   Median (Q1, Q3) | 0.25 (0.08, 0.50) |
|   Range | 0.00 - 2.00 |
| **Hoehn and Yahr scale [at baseline]** | |
|   1 | 169 (42.1%) |

|                                              | Overall (N=401)       |
|----------------------------------------------|-----------------------|
| 2                                            | 232 (57.9%)           |
| **MDS-UPDRS score - sum of parts I-III [at baseline]** |             |
|   N                                          | 401                   |
|   Mean (SD)                                  | 32.78 (13.41)         |
|   Median (Q1, Q3)                            | 31.00 (23.00, 41.00)  |
|   Range                                      | 7.00 - 76.00          |

## 3.2 Analysis of MDS-UPDRS scores in the first 12 months and prior to initiation of symptomatic dopaminergic treatment

A mixed effects model applied to the MDS-UPDRS score in the first 12 months and and prior to initiation of symptomatic dopaminergic treatment gave an intercept (corresponding to the MDS-UPDRS score at baseline) of 33.05 points and a slope of 11.45 points/year. The standard deviations of the linear intercept and slope were 12.21 and 6.60, respectively, with a correlation of 0.58. The residual standard deviation was 6.16.

There was no evidence that any of the covariates gender, age, duation of PD, or Hoehn and Yahr staging scale at baseline affected the slope of progression in MDS-UPDRS scores (overall likelihood ratio test $p=0.34$).

A test for a quadratic rather than a linear mean trajectory was significant ($p=0.02$) and proposed a slight decline in the rate of progression over time though the deviation from linearity was only minor: the linear term in the quadratic model corresponded to an increase by 14.66 points at 12 months whereas the quadratic term corresponded to a -3.66 points lower score at 12 months than predicted by the linear term. Moreover, the quadratic term may be driven by initiation of PD medications other than dopaminergic treatment, because the term was no longer significant ($p=0.49$) if only measurements before any PD medication were included.

**Implications of these analyses for the simulation study**: Longitudinal MDS-UPDRS scores in the control arm were simulated according to a random intercept-slope model. The simulation parameters were chosen based on the PPMI analysis parameters (rounded liberally). ## Timing of initiation of symptomatic dopaminergic treatment

A total of 42.4% of subjects initiated symptomatic dopaminergic treatment within 12 months from baseline. As shown in Table 2, the probability of symptomatic treatment initiation was smaller in the first 6 months than during months 6-12.

Table 2: Timing of initiation of symptomatic dopaminergic treatment.

|                                              | Overall (N=401)       |
|----------------------------------------------|-----------------------|
| **Dopaminergic treatment initiation timing** |                       |
|   Before month 3                             | 7 (1.7%)              |
|   Month 3-6                                  | 20 (5.0%)             |
|   Month 6-9                                  | 102 (25.4%)           |
|   Month 9-12                                 | 41 (10.2%)            |
|   No treatment until month 12                | 231 (57.6%)           |

A marginal logistic model for the probability of symptomatic treatment initiation at each visit depending on the categorical visit and the current UPDRS score gave an odds ratio of 1.48 for each 10 point increase in the UPDRS score. There was no evidence that the change in the MDS-UPDRS score from the previous visit was an additional predictor of dopaminergic treatment initiation ($p$=0.65).

**Implications of these analyses for the simulation study**: In the simulation study, the probability of symptomatic treatment initiation at each visit was simulated according to a logistic regression model depending on the time of the visit (with lower probabilities of initiation during the first 6 months) and the current MDS-UPDRS score (using an OR of 1.5 for each 10 point increase in the MDS-UPDRS score).

### 3.3 Impact of initiation of symptomatic dopaminergic treatment on MDS-UPDRS scores

The difference between the MDS-UPDRS score at the first visit after initiation of dopaminergic treatment compared to the previous assessment (before initiation) is summarized in Table 3.

Table 3: Change in MDS-UPDRS score at the first visit after symptomatic dopaminergic treatment initiation compared to the previous visit.

|  | Overall (N=401) |
|---|---|
| **MDS-UPDRS score change [points]** | |
| N | 127 |
| Mean (SD) | -11.80 (13.15) |
| Median (Q1, Q3) | -10.00 (-21.00, -4.00) |
| Range | -54.00 - 17.00 |

A mixed effects model applied to the MDS-UPDRS score after initiation of dopaminergic treatment in the first 12 months gave a mean slope of 0.96 points/year, i.e. an essentially flat MDS-UPDRS trajectory after initiation of symptomatic dopaminergic treatment.

**Implications of these analyses for the simulation study**: The effect of initiation of symptomatic treatment was simulated according to an immediate improvement in the MDS-UPDRS score with a median drop by 10 points with substantial variability. The fixed slope after initiation of symptomatic treatment was set to 0.

### 3.4 Comparison of "on" and "off" treatment MDS-UPDRS scores

After initiation of symptomatic dopaminergic treatment, 144 MDS-UPDRS scores with part III assessed in a practically defined "off" state ($\geq$ 6 hours after the last dose of dopaminergic treatment) and 251 with part III assessed in an "on" treatment state ($<$ 6 hours after the last dose of dopaminergic treatment) were recorded in our study population in the first year of follow-up. A total of 97 visits with both "on" and "off" treatment values were available. The correlation between "off" and "on" treatment values was high ($\rho$=0.96) and the median difference between "off" and "on" values was 3 (interquartile range 0 to 7) points.

**Implications of these analyses for the simulation study**: As justified in the methods section,

the simulation study was based on simulated "on" treatment MDS-UPDRS scores after initiation of symptomatic dopaminergic treatment. "Off" treatment measurements are slightly larger and less affected by symptomatic treatment but the high correlation between "on" and "off" treatment values suggests that the results of the simulation study are not unduly affected by choosing to model "on" treatment values.

## Acknowledgements

Data used in this appendix were obtained from the Parkinson's Progression Markers Initiative (PPMI) database (www.ppmi-info.org/access-data-specimens/download-data). For up-to-date information on the PPMI study, visit ppmi-info.org. PPMI – a public-private partnership – is funded by the Michael J. Fox Foundation for Parkinson's Research and funding partners listed at www.ppmi-info.org/about-ppmi/who-we-are/study-sponsors.